\documentclass[12pt]{iopart}
\usepackage{graphicx}
\usepackage{epstopdf}
\usepackage{color}
\usepackage{cite}

\renewcommand{\figurename}{Figure}

\begin{document}

\title[Ion heating in Laser interacting with magnetized plasma]
{Ion heating in Laser interacting with magnetized plasma}

\author{Rohit Juneja$^{1*}$, Trishul Dhalia$^1$, Laxman Prasad Goswami$^1$, Srimanta Maity$^2$, Devshree Mandal$^3$ and Amita Das$^{1*}$}

\address{$^{1}$Department of Physics, Indian Institute of Technology Delhi, Hauz Khas, New Delhi 110016, India \\}
\address{$^{2}$ Extreme Light Infrastructure ERIC, ELI Beamlines Facility, Za Radnicí 835, 25241 Dolní Břežany, Czech Republic \\}
\address{$^{3}$ Institute for Plasma Research, Gandhinagar, Gujarat, 382428, India}
\ead{$^{*}$onlyforjuneja@gmail.com and amita@iitd.ac.in}
\vspace{10pt}

\begin{abstract}
   The ion heating mechanism in the context of laser interacting with plasma immersed in a strong magnetic field is studied. The magnetic field is chosen to be strong for laser electromagnetic field propagation inside the plasma to be governed by the magnetized dispersion relation. Both  X and RL mode configurations have been studied in detail using Particle - In - Cell (PIC) simulations. It is shown that the energy absorption process is governed by a resonant mechanism wherein the laser frequency matches with an underlying mode in the plasma. For X  and RL mode configurations, these correspond to lower hybrid and ion cyclotron resonance, respectively. The absorption, however, is found to be most efficient at frequencies close to but not exactly matching with the resonance frequency. An understanding of the same has been provided. The role of laser polarization has been studied in detail.\\
   
\end{abstract}

\section{Introduction}\label{sec:Introduction}
Ion heating is one of the key research topics in the context of laser-plasma interactions. There are many applications  where it is desirable to have energetic ions \cite{borghesi2010progress, roth2001fast,  atzeni2002first, bulanov2002feasibility, naumova2009hole, steward1973proton}. There are various mechanisms by which plasma heating can be obtained by a laser pulse \cite{kaw1969laser,kaw2017nonlinear,das2020laser}. 
In the un-magnetized laser-plasma interaction, the laser energy gets coupled with  the lighter 
electron species through various collisional and collisionless mechanisms \cite{Kruer1985JBHB, Brunel1987NotsoresonantRA, ping2008absorption,  wilks1992absorption, chopineau2019identification, freidberg1972resonant, stix1965radiation, gibbon1992collisionless}. The collisionless schemes are either based on  resonant processes (in which the laser frequency matches with the plasma frequency) or rely on the presence of a component of laser electric field and/or $\vec{J} \times \vec{B}$ force acting perpendicular to the plasma surface which  pulls the electrons outside in the vacuum region. There are efficient mechanisms by which ion heating and/or acceleration  can be obtained \cite{esirkepov2004highly, kumar2019excitation, turrell2015ultrafast, pfotenhauer2010cascaded, macchi2013ion, henig2009radiation, robson2007scaling, snavely2000intense, fuchs2006laser, robinson2008radiation, haberberger2012collisionless, wilks2001energetic}. The application  of an external magnetic field on a plasma target  has several  advantages. The laser can propagate  even inside an overdense plasma target at the various pass bands permitted by the magnetized plasma dispersion relation. There are many modes in the magnetized plasma that can resonate with the laser frequency. Each of these modes would   have its own characteristic mechanism  of contributing to the heating/coupling processes with the plasma species. Furthermore, 
if  the laser frequency lies between electron and ion gyro-frequencies, the motion of electrons 
gets  constrained  by the magnetic field. In such a scenario, the heavier ions species have a dominant dynamical response to the laser field.   Therefore, in this regime,  the 
ions can couple directly with the laser field and acquire energy from it. 

 The laser-plasma interaction studies have  so far been  primarily confined to the domain of  un-magnetized plasma response. This is primarily because the  magnetic field requirements turn out to be quite high, even 
 to make the lighter 
 electron species  magnetized  at the laser frequency. Recent developments in producing high magnetic fields of the order of kilo Tesla \cite{nakamura2018record, korneev2015gigagauss} would typically elicit a  magnetized response from the lighter electron species 
  at the low frequency of $CO_2$ lasers. This has opened up the possibility of studying  the regime of laser interacting with magnetized plasmas. Recently, a number of studies have been carried out to explore this particular regime using Particle - In - Cell (PIC) simulations \cite{vashistha2021excitation, maity2022mode, vashistha2022localized, goswami2022observations}. New  mechanisms of laser energy coupling to ions 
  have been proposed, which rely on generating electrostatic fields as a result of 
  the difference between  $\vec{E} \times \vec{B}_0$ (Here $\vec{E}$ is the oscillating electric field of the laser and $\vec{B}_0$ is the applied magnetic  field) drifts and/or ponderomotive force felt by ions and electrons \cite{vashistha2020new, goswami2021ponderomotive}. The  generation of  harmonic radiation was shown to be efficient at a certain optimal value of the external magnetic field \cite{maity2021harmonic}. Another interesting result was the observation of complete transparency of the plasma at a very high magnetic field \cite{mandal2021electromagnetic}.
  
  In this study, using Particle-In-Cell (PIC) simulations, we have explored different mechanisms of 
laser energy absorption 
relying on the laser frequency being resonant with the  magnetized plasma modes. In particular, we have chosen the lower hybrid resonance in the context of the X-mode configuration and the ion cyclotron resonance in the R-L mode configuration of the applied magnetic field. It is shown that the  absorption of laser energy is best when  the laser frequency is slightly off-resonance. The dependence of energy absorption on laser polarization has also been investigated.

This paper contains the following sections. In section \ref{sim}, simulation details have been presented. 
 In section \ref{sec4}, the X-mode configuration with applied magnetic field directed along laser magnetic field direction  has  been chosen. 
The absorption near the LH resonance has  been studied in detail. It is shown that peak absorption occurs when the laser frequency is slightly lower than the lower hybrid  resonant frequency. An understanding of the same has been provided. 
The observations and analysis for the  case in which the applied magnetic field is along the laser propagation direction, i.e. the RL mode geometry has been presented in section \ref{sec3}. Here, we present a detailed study on the energy absorption process when the laser frequency is close to ion gyro-frequency. 
It is shown that there are two distinct processes of laser energy absorption by ions. In one case, the transfer is mediated by electrostatic fluctuations generated in the medium. However, in the second case of ion gyro-resonance, the electrostatic excitations are altogether missing. 
In section \ref{sec5}, we discuss the role of laser polarization on the absorption process in each of these configurations. Section \ref{sec7} contains the conclusion of the work. 

\begin{figure}
  \centering
  \includegraphics[height=2in,width=6.0in]{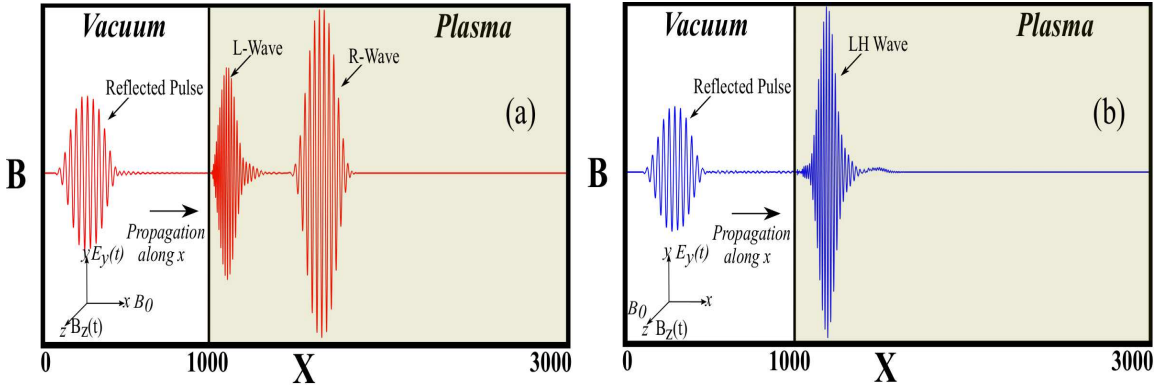}
  \caption{Schematic(not to scale) showing that in (a) Laser frequency lies in the pass band of RL Mode configuration where both R and L waves are obtained in the plasma and in (b) Laser frequency lies in the first pass band of X-mode where LH wave is obtained in the plasma.}
\label{figure1}
\end{figure}

\section{Simulation details}
\label{sim}

In this study, one-dimensional (1D) Particle-In-Cell (PIC) simulations have been performed using a massively parallel particle-in-cell code OSIRIS 4.0 \cite{hemker2000particle, fonseca2002osiris, fonseca2008one}. A 1D simulation box with length $L_x = 3000 c/\omega_{pe}$ has been considered. Here, $c$ is the speed of light in vacuum, $\omega_{pe}$ represents the electron-plasma frequency. In all our simulations the 
 time has been normalised by  $t_{n}$ = ${\omega_{pe}}^{-1}$, for length, the skin depth  $x_{n} = c/\omega_{pe}$ has been chosen. The electric and magnetic fields are normalized by $E_{n} = B_{n} = m_{e}c\omega_{pe}/e$, where $m_{e}$ is the mass of the electron and $e$ is the magnitude of the  electronic charge.  Absorbing boundary conditions have been used for both particles and fields in both directions. We have considered $60000$ grid points (cells) in our simulations, corresponding to the grid size $dx = 0.05c/\omega_{pe}$. The temporal resolution is chosen to be as $dt = 0.02\omega_{pe}^{-1}$.  The number of particles per cell is taken to be 8. From $x=0$ to $1000c/\omega_{pe}$, there is a vacuum region, and the plasma boundary starts from  $1000c/\omega_{pe}$.

The schematic of the simulation geometry is shown in Fig.(\ref{figure1}). The direction of propagation of the laser pulse is in the $\hat{x}$-direction. The electric and magnetic fields component of the laser pulse is in $\hat{y}$ and $\hat{z}$-direction, respectively. We have also considered cases in which the electric field is circularly polarised. Both  right and left hand circular polarizations have been considered. In the former case,  the electric field vector rotates from $E_y$ to $E_z$, while it rotates vice-versa for left circular polarization. In section \ref{sec4} the applied magnetic field is chosen to be directed along $\hat{z}$, the same direction as that of the laser magnetic field for linearly polarised laser. It should be noted that for the choice of circular polarisation discussed in \ref{sec5} this configuration leads to a combination of O- and X-modes. 
In section \ref{sec3}, we study the case in which the applied magnetic field ($B_0$) is along $\hat{x}$ the laser propagation direction. We also call this as the 
 RL-mode configuration as the normal modes of the plasma in this configuration satisfies the $R$ and $L$ mode dispersion relations.   Table 1 contains laser and plasma parameters in normalized units and corresponding values in the standard units. Ion mass is taken to be $25$ times to  the mass of the electrons for faster computational calculations. For some parts of this study, a higher mass ratio has also been chosen to validate the observations. A laser pulse with intensity $3.85 \times 10^{15} W cm^{-2}$ (corresponding to the normalised vector  potential, $a_{0}=eE_{l}/m\omega_{l}c=0.5$) has been chosen for linearly polarized laser pulse.  For comparison with circular polarisation studies $a_{0}=0.5/\sqrt{2}$ has been  considered  to ensure that the   total energy of the incident laser for the linear and circularly polarised cases   are identical. The laser frequency in all the simulations has been fixed at   $ 0.2\omega_{pe}$. The magnetic field is varied to study different configurations.

\begin{table}
    \caption{Simulation parameters in normalized units and possible values in standard units.} 
  \begin{center}

  \begin{tabular}{lcc}
  \hline
      \textbf{Parameters}  & \textbf{Normalised value}   &   \textbf{Value in standard units}\\
      \hline \\
       & \textbf{Laser Parameters}  & \\
      Frequency ($\omega_{L}$)  & $0.2 \omega_{pe} $ & $ 0.2 \times 10^{15} Hz$\\
      Wavelength ($\lambda_{L}$)  & 31.4$c/\omega_{pe}$ &9.42 $\mu m$\\
      Intensity  & $a_0 = 0.5$ & $3.85\times10^{15}Wcm^{-2}$\\\\
       & \textbf{Plasma Parameters} &\\
      Number density ($n_{0}$) & 1 & $3.15 \times 10^{20} cm^{-3}$\\
       Electron Plasma Frequency\\   ($\omega_{pe}$) & 1 & $10^{15} Hz$\\
      \hline
  \end{tabular}
  \end{center}
  \label{Table}
\end{table}

\section{The configuration of X mode}
\label{sec4}
\begin{figure}
  \centering
  \includegraphics[width=6.0in]{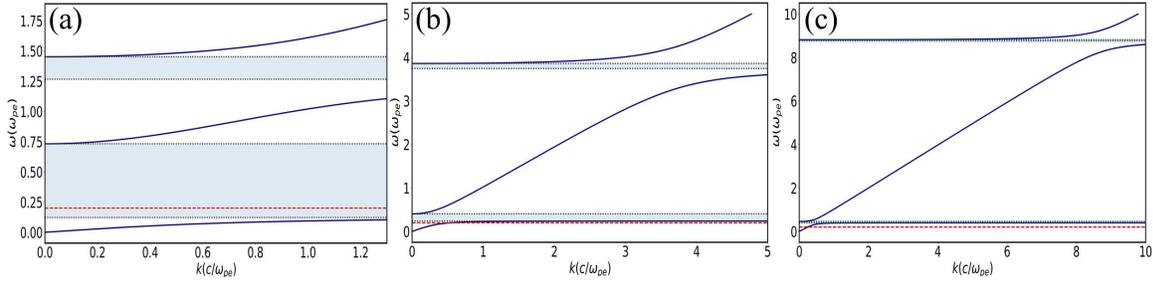}
  \caption{Dispersion curves in X mode geometry showing pass bands and stop bands for various $\omega_{lh}$ values. (a) $\omega_{lh}$=0.12$\omega_{pe}$, (b) $\omega_{lh}$=0.24$\omega_{pe}$, and (c) $\omega_{lh}$=0.40$\omega_{pe}$ with corresponding external magnetic fields $B_{0} = 0.75,$ $3.6,$ $and$ $8.7$ respectively.}
\label{figure9}
\end{figure}

We now consider the case in which the applied magnetic field is aligned perpendicular to the laser propagation direction in the X-mode configuration. The dispersion relation for the wave has been shown 
in Fig.(\ref{figure9}) for various strengths of the magnetic field that corresponds to different Lower Hybrid frequencies $\omega_{lh} = 0.12 \omega_{pe}$, $0.24 \omega_{pe}$ and $0.4 \omega_{pe}$. The red dotted line corresponds to our choice of laser frequency ($\omega = 0.2 \omega_{pe}$). For the first case (Case I) the lower hybrid frequency is $\omega_{lh} = 0.12 \omega_{pe}$ and hence the laser frequency lies inside the stop band. For the other  cases,  (II) $\omega_{lh} = 0.24 \omega_{pe}$ and  (III) $\omega_{lh} = 0.4 \omega_{pe}$ the laser frequency lies in the pass band of X mode. However, for Case(II) the laser frequency is comparatively nearer to  the lower hybrid resonance than in Case(III). For studies considered in this section we have considered only the linear polarisation of incident laser radiation.

\begin{figure}
  \centering
  \includegraphics[width=6.0in]{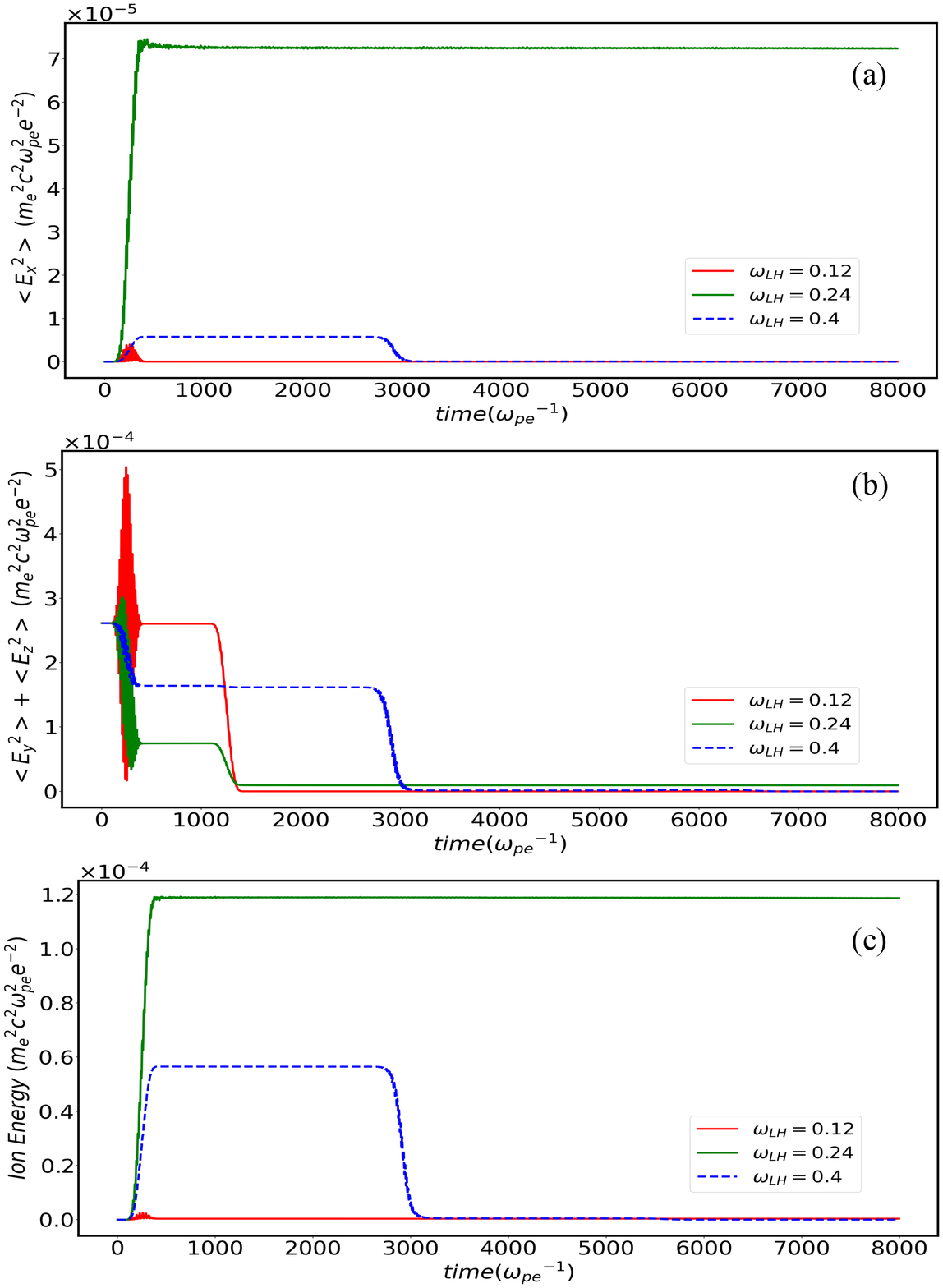}
  \caption{Time evolution of (a) electrostatic energy, (b) electromagnetic energy, and (c) kinetic energy of ions in X Mode configuration.}
\label{figure11}
\end{figure}

 \begin{figure}
  \centering
  \includegraphics[height=2.5in,width=6.0in]{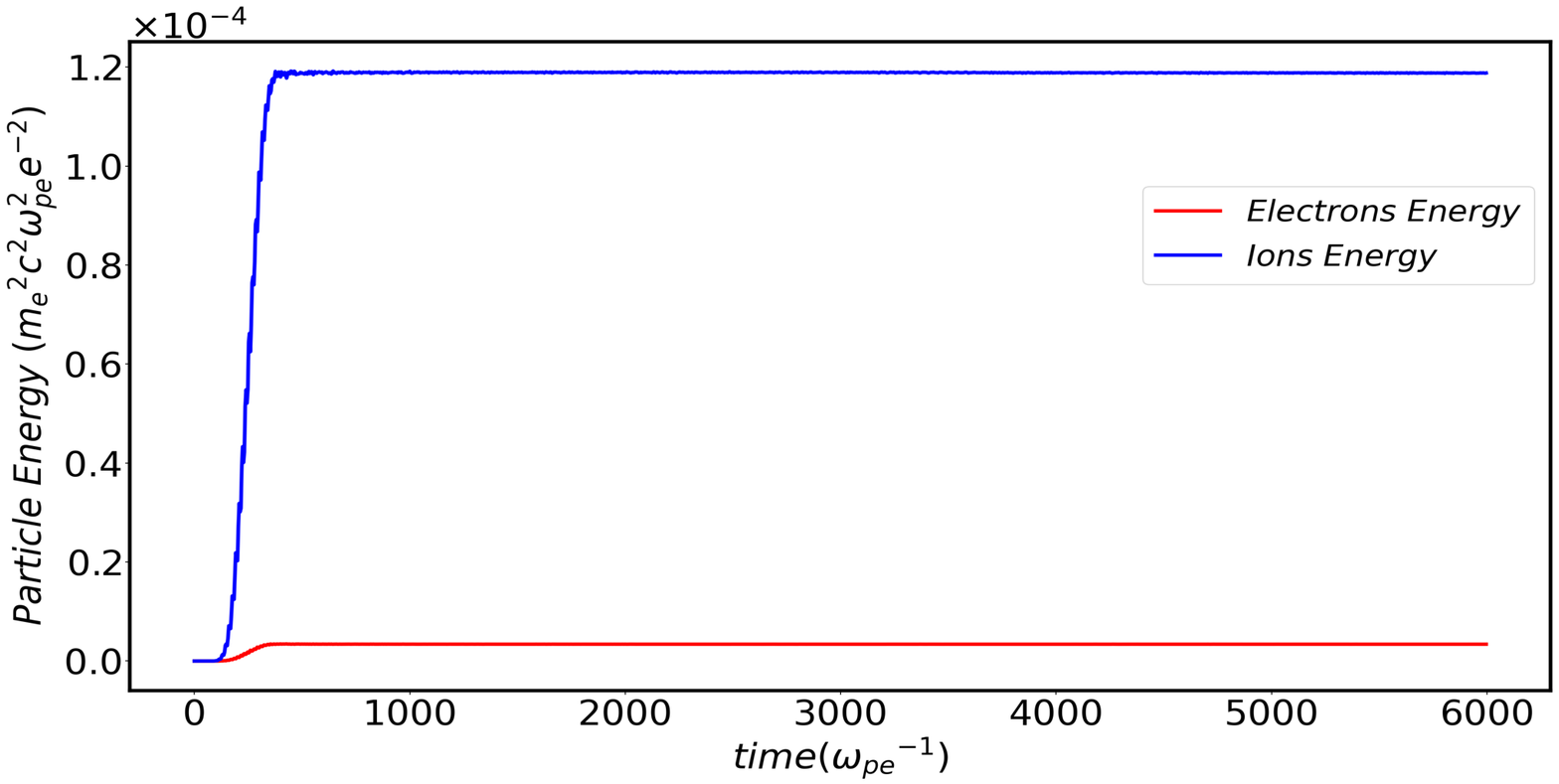}
  \caption{Comparison between energy absorbed by electrons and ions for X-mode configuration for Case(II).}
\label{Fig:Xmode-el+ion-compare}
\end{figure}

\begin{figure}
  \centering
  \includegraphics[height=5in,width=6.0in]{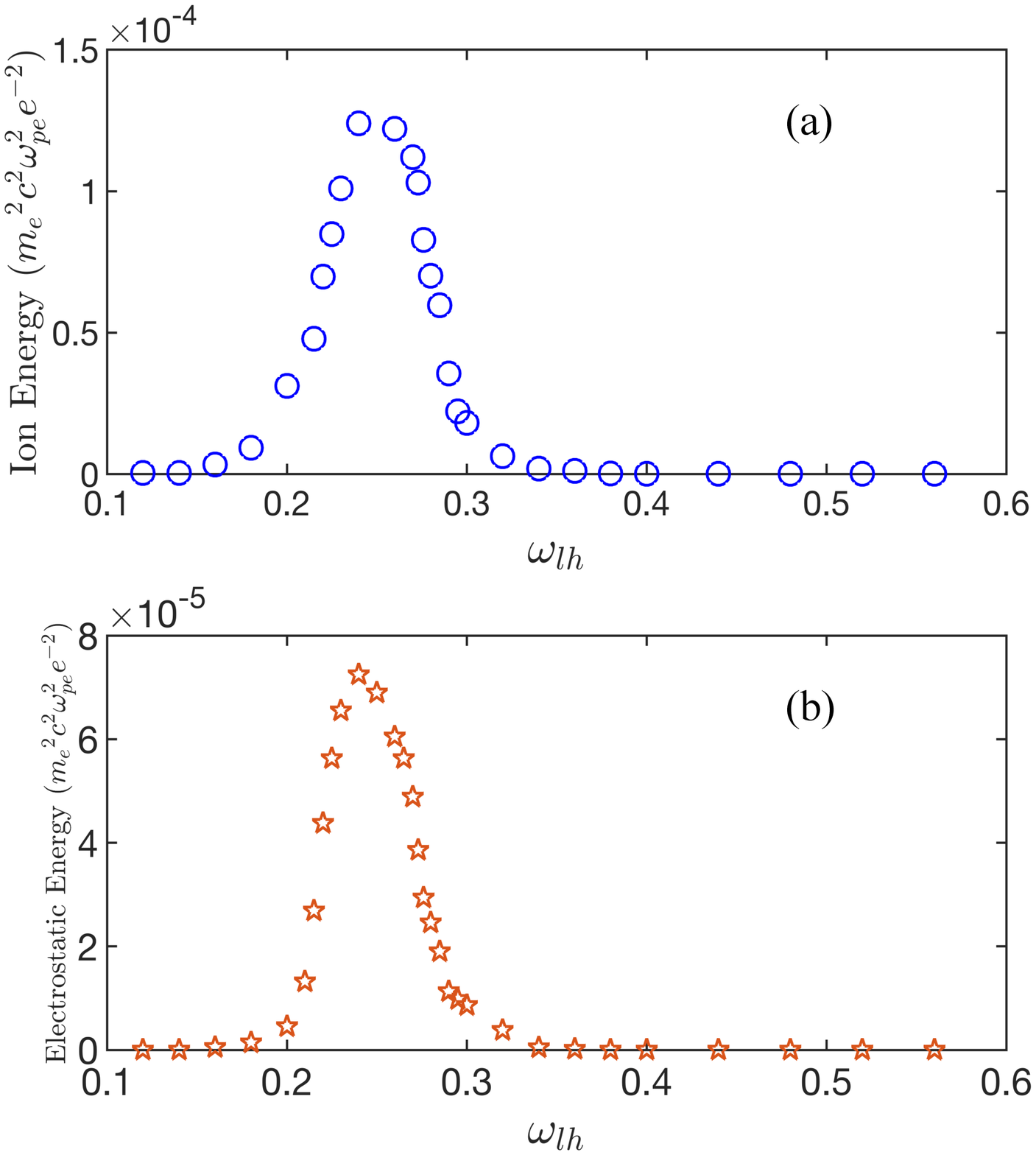}
  \caption{Variation of (a) ion energy, and (b) electrostatic energy with $\omega_{lh}$.}
\label{figure10}
\end{figure}

A comparison of the three cases has been shown in Fig.(\ref{figure11}). In this figure, we  have shown for all three cases the time evolution of the electrostatic energy (which gets generated as a result of the interaction of the laser with the plasma), the electromagnetic energy, and the particle kinetic energy. It can be observed from the figure that the electrostatic and particle energy increases with time, whereas the electromagnetic energy associated with the incident wave decreases. The  sharp dips in the electromagnetic energy evolution indicate the time when either the  reflected wave  (from the plasma surface) or the transmitted wave moves  outside the simulation box. For Case(I) the laser frequency being in the stop band the interaction of the laser with plasma occurs only over a small skin depth region. It can be observed that in this case, the electrostatic, as well as the particle kinetic energy, is negligible. For Case(II), where the incident laser frequency is near lower hybrid resonance, both these energies are finite and remain so even after the laser has left the simulation box. For Case(III) though the electrostatic and particle energy acquires some finite value during the time laser remains in the simulation box, they ultimately become negligible, indicating that the process of energy transfer was reversible in this case. We next provide a comparison of the time evolution of  kinetic energy of ions and electrons for Case(II)  in Fig.(\ref{Fig:Xmode-el+ion-compare}).  It  can be observed that energy absorbed by ions is much higher as compared to electrons for this case.
\begin{figure}
  \centering
  \includegraphics[width=6.0in]{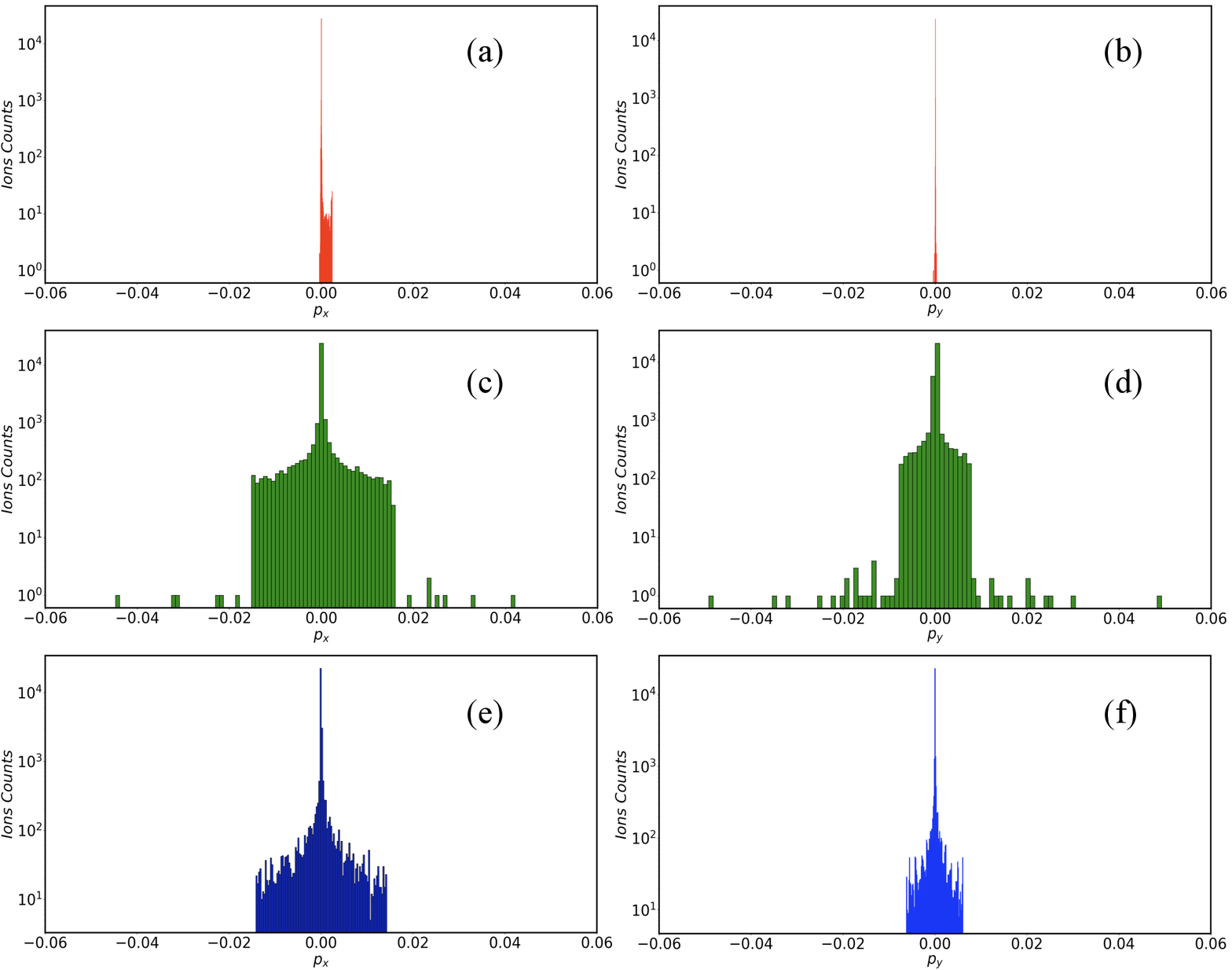}
  \caption{Longitudinal momentum distribution in (a), (c) and (e) and Transverse momentum distribution in (b), (d) and (f) for $\omega_{lh} = 0.12 \omega_{pe}$ (Red Color Plots), $\omega_{lh} = 0.24 \omega_{pe}$ (Green Color Plots) and $\omega_{lh} = 0.4 \omega_{pe}$ (Blue Color Plots), respectively, in X Mode configuration at $t = 2000$ when the laser is still in the simulation box.}
  \label{Fig:ion-counts-xmode-2000}
\end{figure}

\begin{figure}
  \centering
  \includegraphics[width=6.0in]{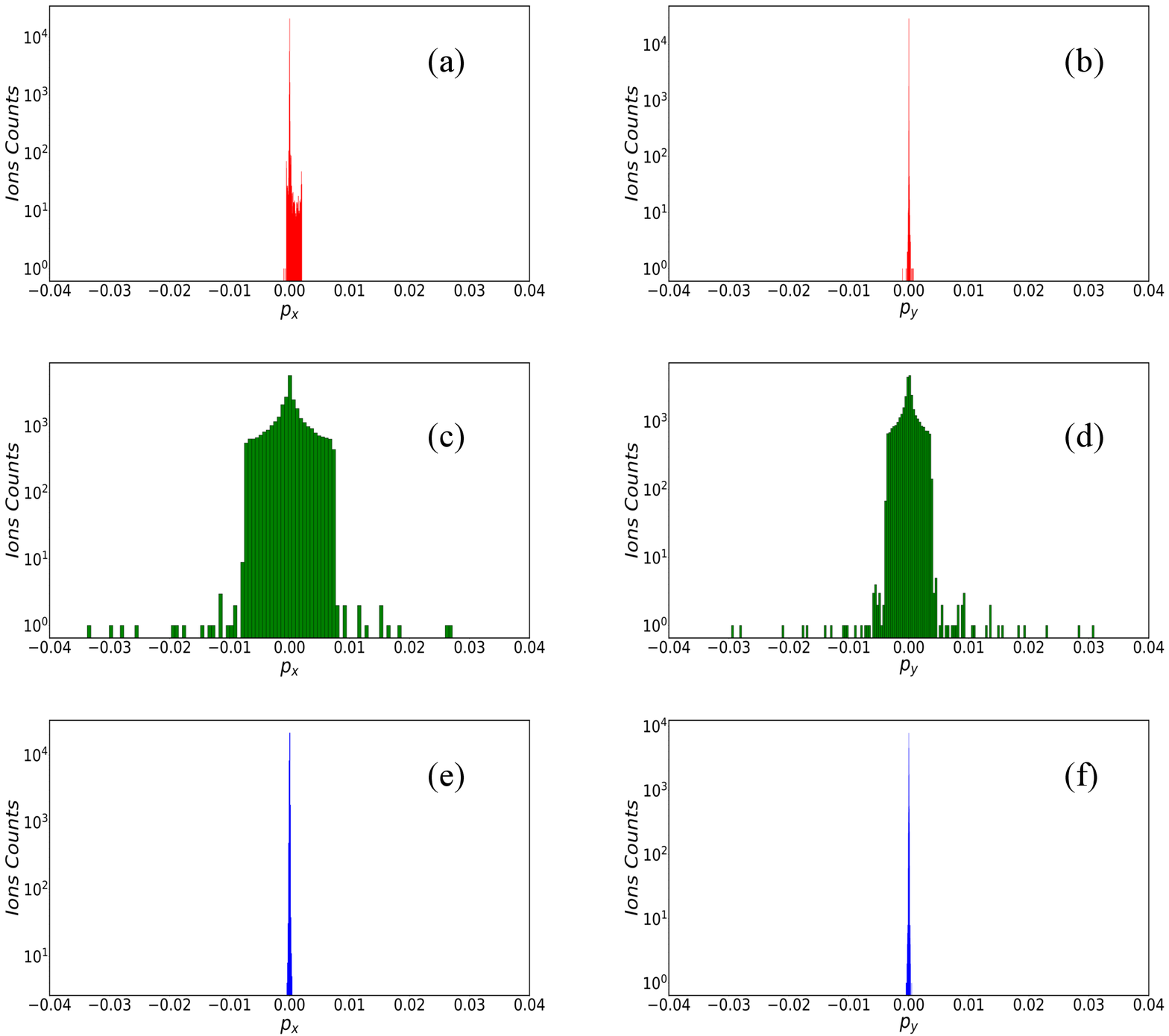}
  \caption{Longitudinal momentum distribution in (a), (c) and (e) and Transverse momentum distribution in (b), (d) and (f) for $\omega_{lh} = 0.12 \omega_{pe}$ (Red Color Plots), $\omega_{lh} = 0.24 \omega_{pe}$ (Green Color Plots) and $\omega_{lh} = 0.4 \omega_{pe}$ (Blue Color Plots), respectively, in X Mode configuration at $t = 7800$ when laser has left the simulation box.}
  \label{figure12}
\end{figure} 

\begin{figure}
  \centering
  \includegraphics[height=2.5in,width=6.0in]{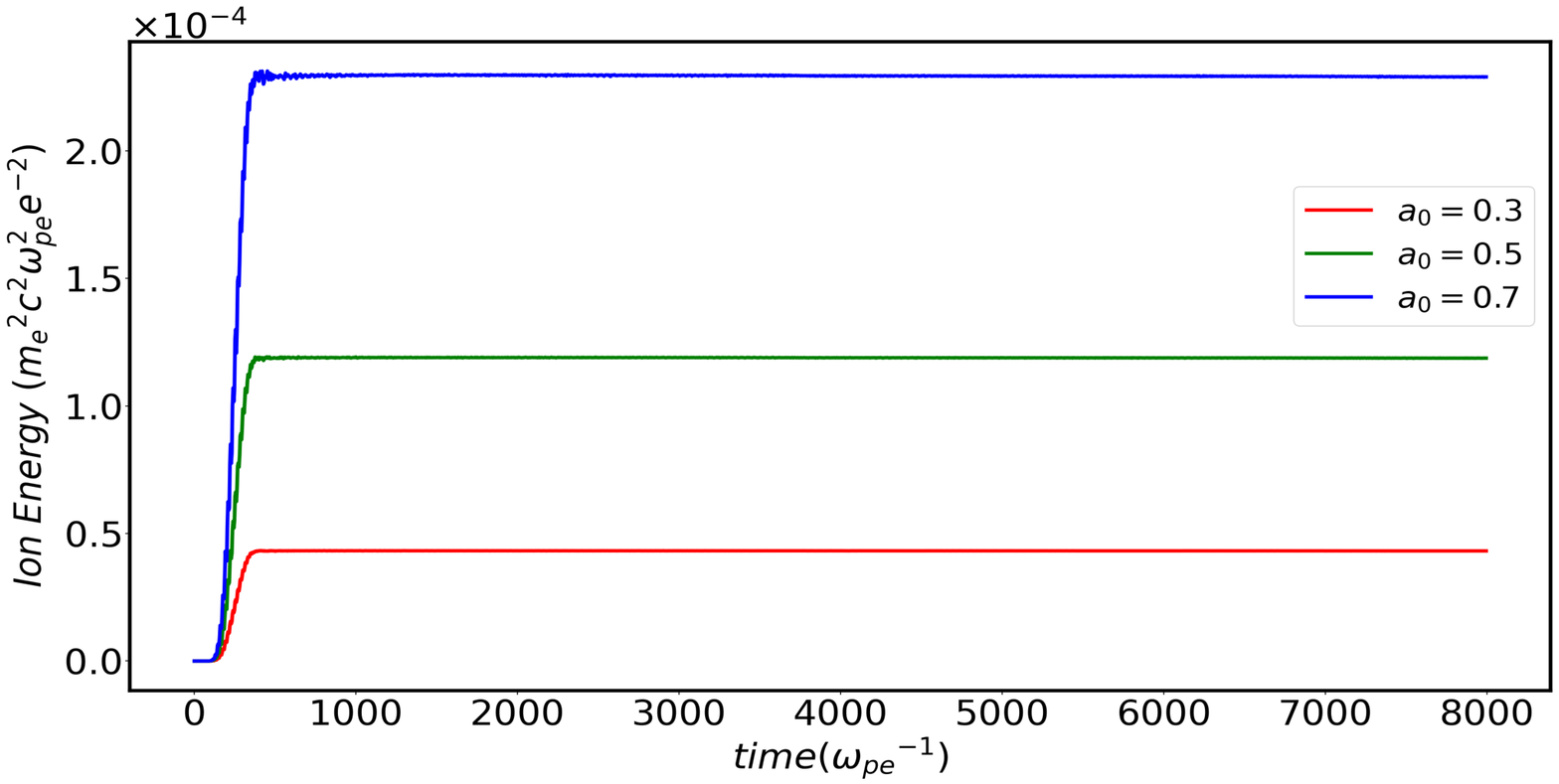}
  \caption{Comparison of energy absorbed by ions as a function of laser intensity in X-mode geometry for Case(II).}
\label{figure14}
\end{figure}

We have also carried out simulations for different magnetic fields for which the laser frequency  (chosen as $0.2 \omega_{pe}$) lies in the  stop band to  deep inside the pass band of the dispersion curve. The energy acquired by the ion has been shown in subplot(a) of  
 Fig.(\ref{figure10}) as a function of $\omega_{lh}$. In subplot(b) of the same figure, the electrostatic field energy generated in the system  has been shown.  These energies correspond to the time  after the laser pulse has left the system.  

We  observe that  by  increasing  $\omega_{lh}$  (i.e., choosing a higher external magnetic field, $B_{0}$), the energy absorption at first increases, reaches a peak, and then subsequently starts to decrease. The peak of energy absorption for electrostatic energy occurs  around $\omega_{lh}$=0.24$\omega_{pe}$ as shown in subplot(b) of Fig.(\ref{figure10}).  
The peak of ion energy absorption does not occur at the exact lower hybrid frequency. It occurs when the laser frequency is slightly lower than $\omega_{lh}$.   The decay on both sides can be understood by realizing that absorption depends on the interplay of two effects.  The absorption will be high  if the laser pulse gets sufficient time to interact with  the plasma. If the laser frequency   is in the stop band then the absorption gets  limited only to a very small region  of the order of skin depth where it penetrates  the plasma medium. If the  laser frequency is higher than   $\omega_{lh}$ it lies in the stop band and the larger the difference between the two frequencies smaller the penetration depth, thereby limiting the energy absorption. On the other hand, when the laser frequency is smaller  than the $\omega_{lh}$ it lies in the pass band of the X mode dispersion curve and hence propagates freely inside the plasma. In this case, however, the efficiency of absorption gets decided by how close it is to the Lower hybrid resonance frequency to excite an electrostatic mode from which the energy is subsequently taken over by the ions. Here the energy absorption is mediated by the resonant lower hybrid electrostatic mode. Hence the   electrostatic energy  typically displays the same trend with magnetic field variation as  shown by the  ion energy gain.  

We now show the ion counts for the momenta along the laser propagation and the perpendicular direction for two different times of $t = 2000$ and $ t= 7800$ respectively in Fig.(\ref{Fig:ion-counts-xmode-2000}) and Fig.(\ref{figure12}) respectively. At $t = 2000$ the laser pulse is in the system, and at $t= 7800$ it has left the system. 
The ion momenta along both propagation and perpendicular direction have a considerable width only for Case(II) 
for which the energy absorption is maximum. 

The effect of laser intensity has been shown in Fig.(\ref{figure14}), where we have taken three different values of $a_{0}$ for this configuration. The  Red, green, and blue curves represent $a_{0}=0.3$, $0.5$, and $0.7$, respectively. As expected,  with increasing  laser intensity, energy absorbed by ions increases.

\section{The configuration of RL mode}
\label{sec3}

\begin{figure}
  \centering
  \includegraphics[width=6.0in]{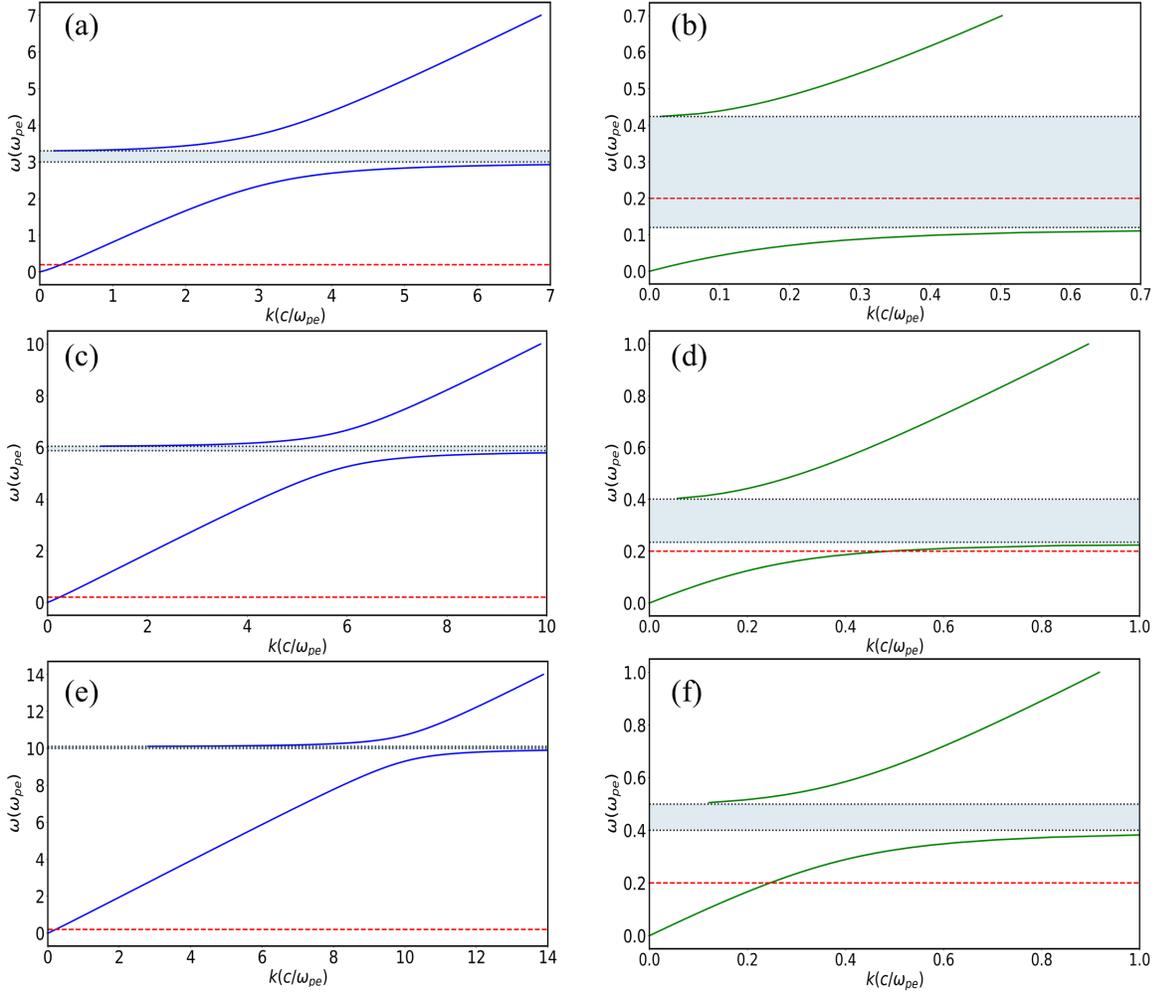}
  \caption{Dispersion curves in RL mode geometry showing pass bands and stop bands for various $\omega_{ci}$ values. (a),(c),(e) showing R mode dispersion curves for $\omega_{ci}= 0.12\omega_{pe}$, $0.235\omega_{pe}$, $0.40\omega_{pe}$  respectively and (b),(d),(f) showing L mode dispersion curves for $\omega_{ci}$= 0.12$\omega_{pe}$, 0.235$\omega_{pe}$, 0.40$\omega_{pe}$ values respectively with corresponding external magnetic fields $B_{0} = 3,$ $5.875,$ $and$ $10$ respectively.}
\label{figure5}
\end{figure}

In this section, we discuss the laser interaction with plasma  for a case where the external magnetic field is  parallel to the laser propagation direction, i.e., along $\hat x$. This configuration is termed as the R-L mode configuration as the right circular polarized (R-mode) and left circular polarized (L-mode) are the permitted normal modes in the plasma in this configuration.  The dispersion curves of both $R$ and $L$ modes with  their respective stop and pass bands for the  frequency  of electromagnetic wave  have been shown in Fig.(\ref{figure5}). The subplots in various rows  are for  three different values of the applied magnetic field, which correspond to the ion gyro-frequency of $ \omega_{ci} = 0.12$, $ 0.235$, and $0.40$ in the normalised units.
With increasing magnetic field, the region of the stop band diminishes. 
We consider the incident laser to be linearly polarised for the simulation studies presented in this section.

\begin{figure}
  \centering
  \includegraphics[width=6.0in]{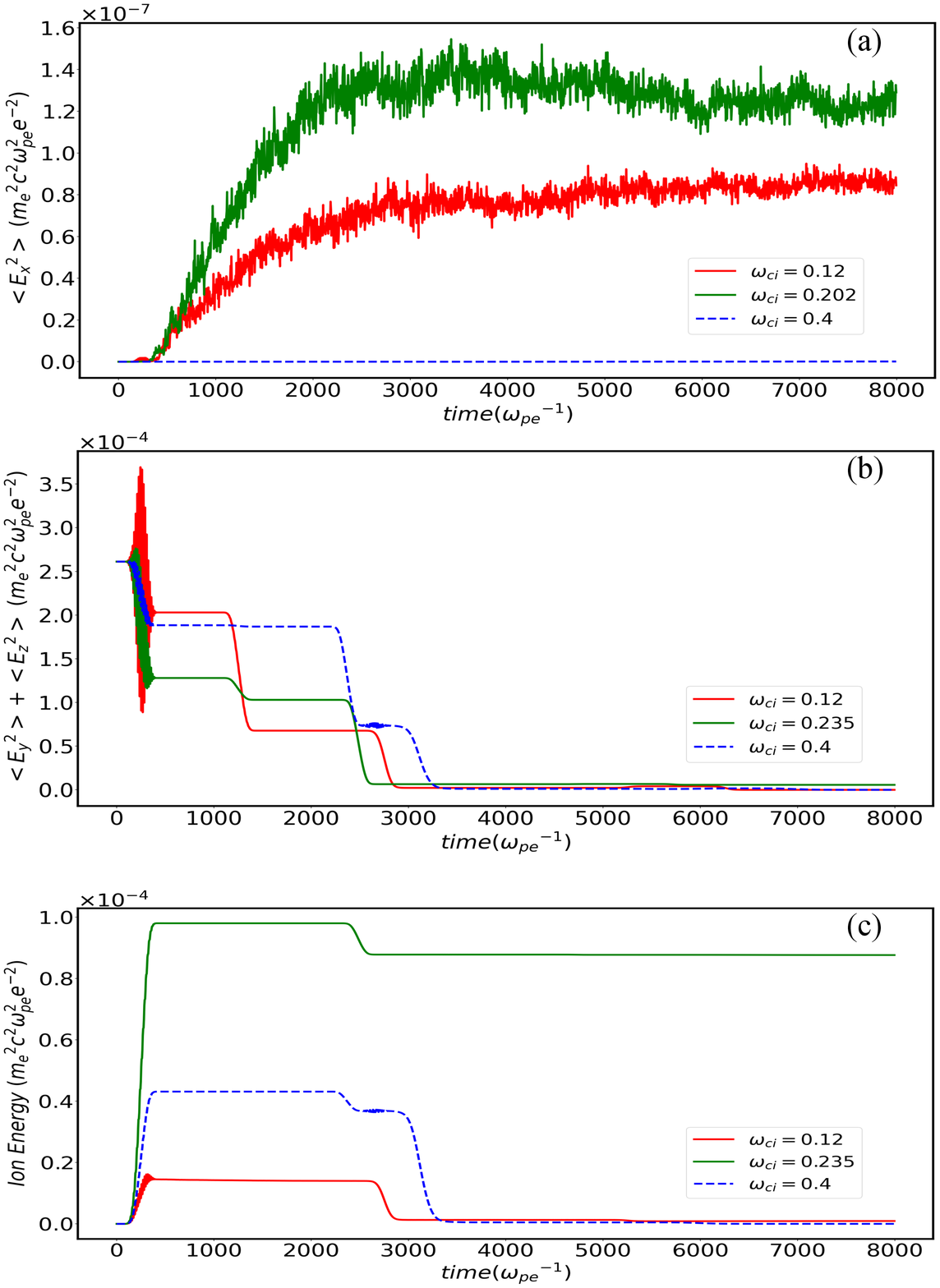}
  \caption{Time evolution of (a) electrostatic energy, (b) electromagnetic energy, and (c) kinetic energy of ions in RL Mode configuration.}
\label{figure7}
\end{figure}

\begin{figure}
  \centering
  \includegraphics[height=2.5in,width=6.0in]{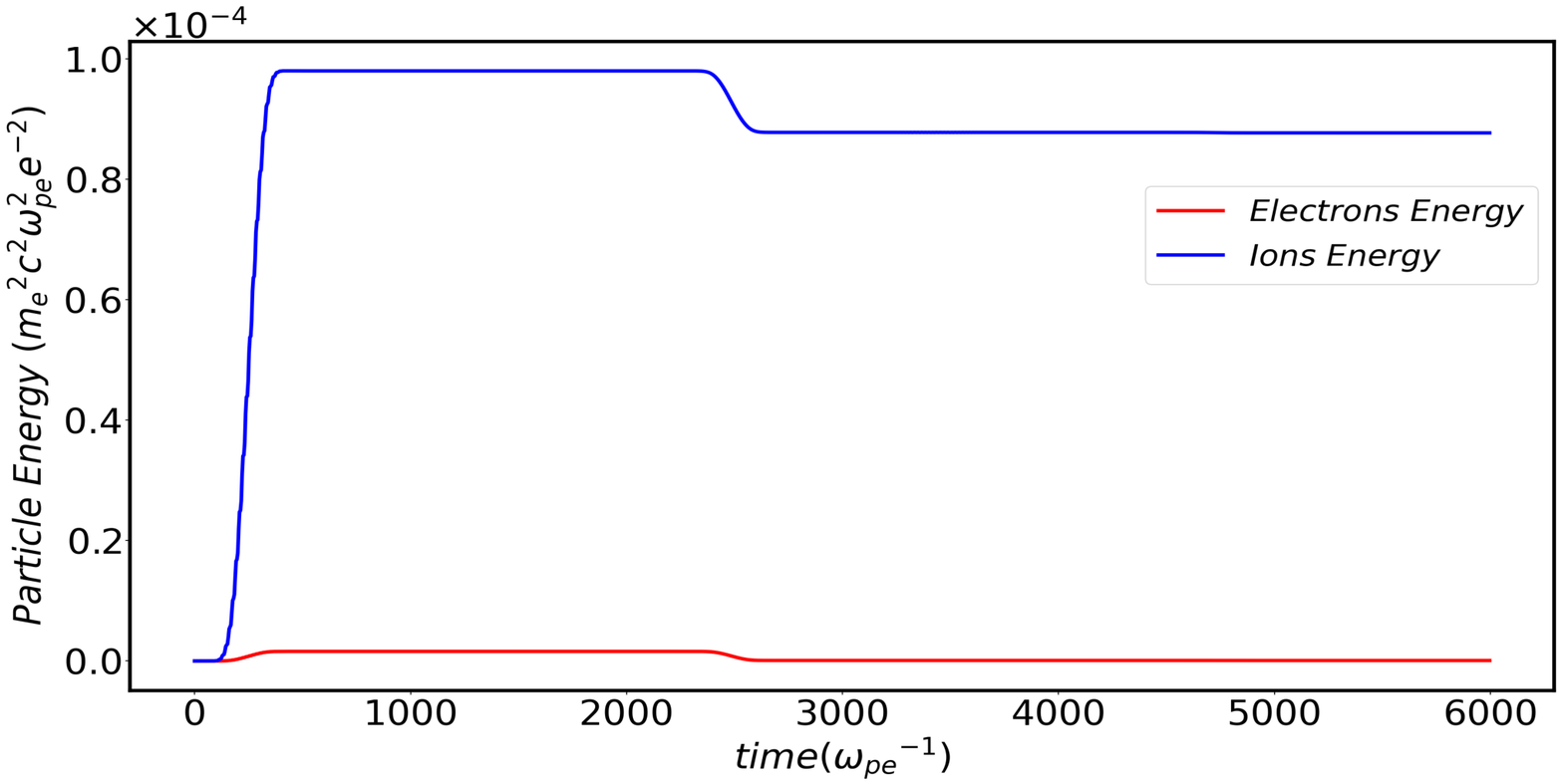}
  \caption{Comparison between energy absorbed by electrons and ions for R-L mode configuration for Case(II).}
\label{Fig:RLmode-el+ion-compare}
\end{figure}

\begin{figure}
  \centering
  \includegraphics[height=5in,width=6.0in]{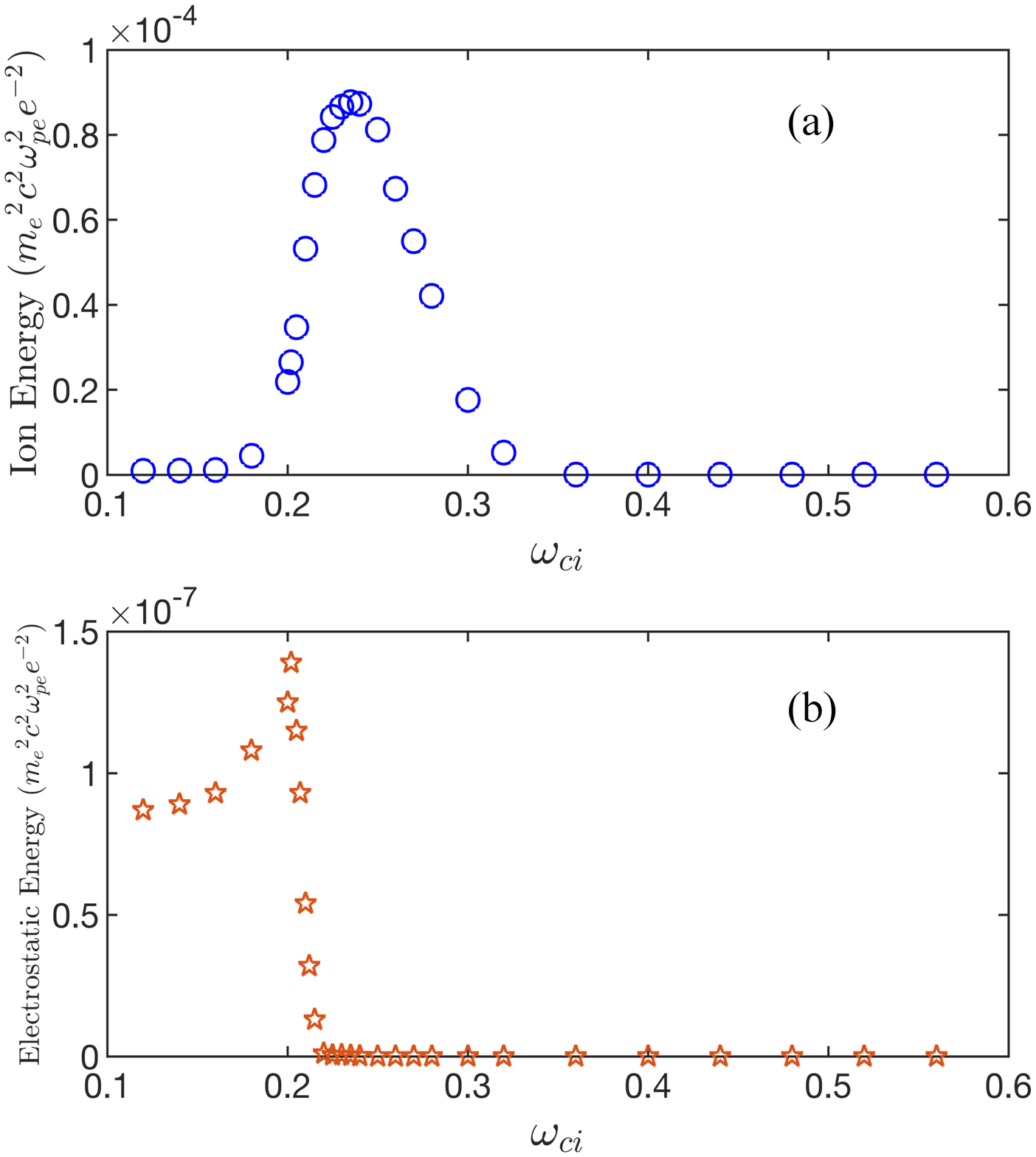}
  \caption{Variation of (a) ion energy, and (b) electrostatic energy with $\omega_{ci}$.}
\label{figure6}
\end{figure}

 \begin{figure}
  \centering
  \includegraphics[width=6.0in]{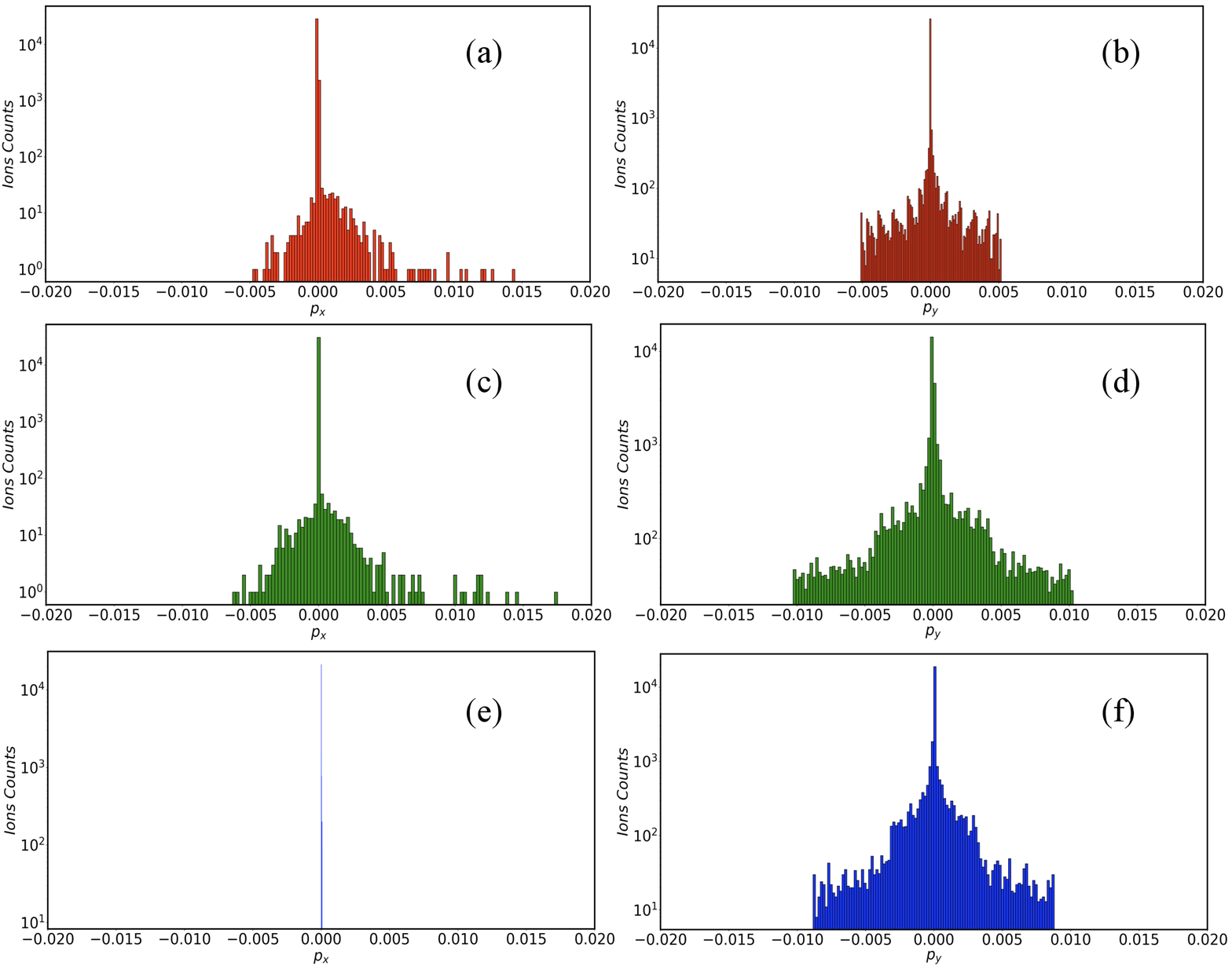}
  \caption{Longitudinal momentum distribution in (a), (c) and (e) and Transverse momentum distribution in (b), (d) and (f) for Case I (Red Color Plots), II (Green Color Plots) and III (Blue Color Plots) respectively in RL Mode configuration at $t = 2000$ when the laser is still in the simulation box.}
\label{figure19}
\end{figure}

\begin{figure}
  \centering
  \includegraphics[width=6.0in]{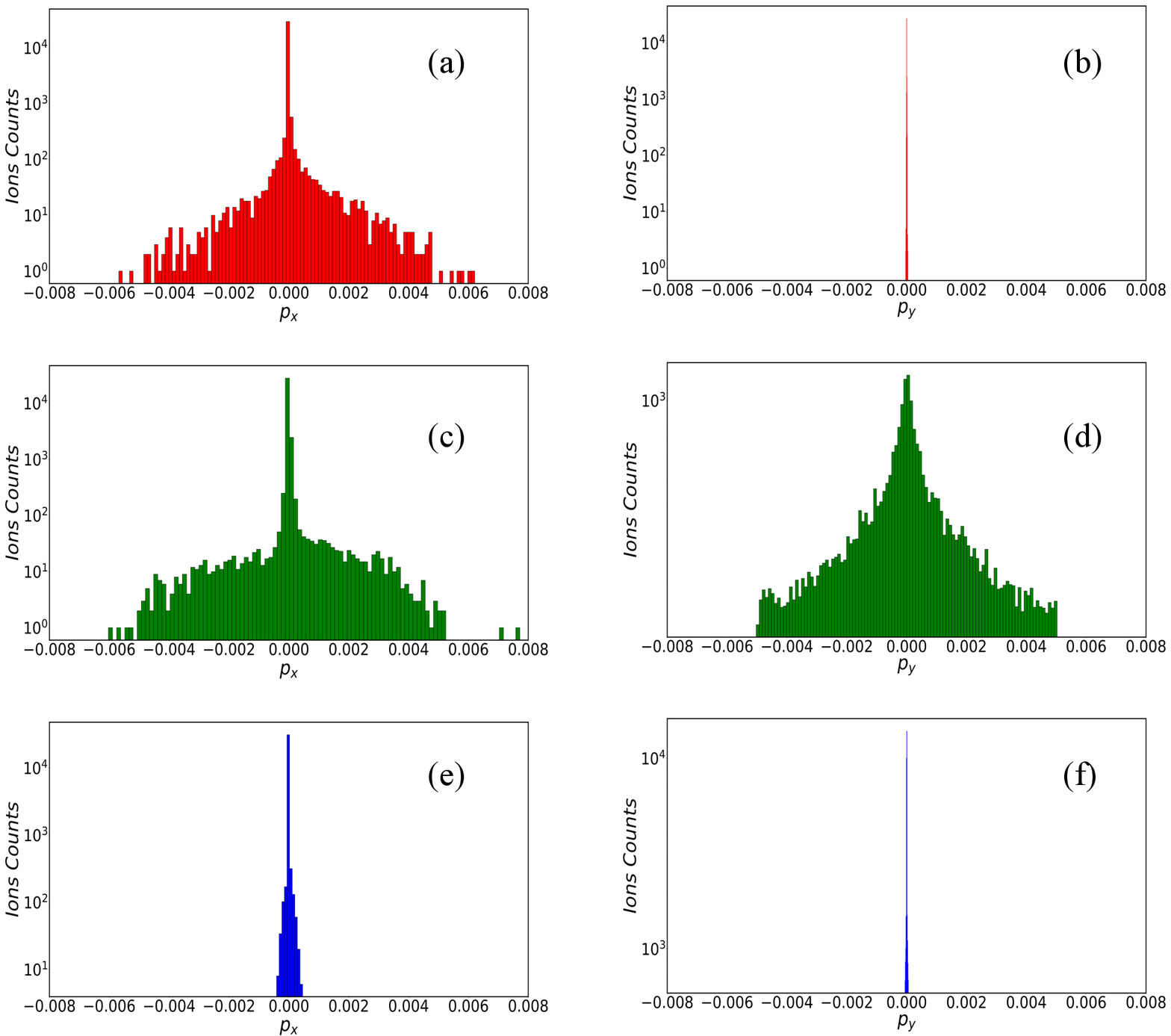}
  \caption{Longitudinal momentum distribution in (a), (c) and (e) and Transverse momentum distribution in (b), (d) and (f) for Case I (Red Color Plots), II (Green Color Plots) and III (Blue Color Plots) respectively in RL Mode configuration at $t = 7800$ when laser has left the simulation box.}
\label{figure8}
\end{figure}

\begin{figure}
  \centering
  \includegraphics[height=2.5in,width=6.0in]{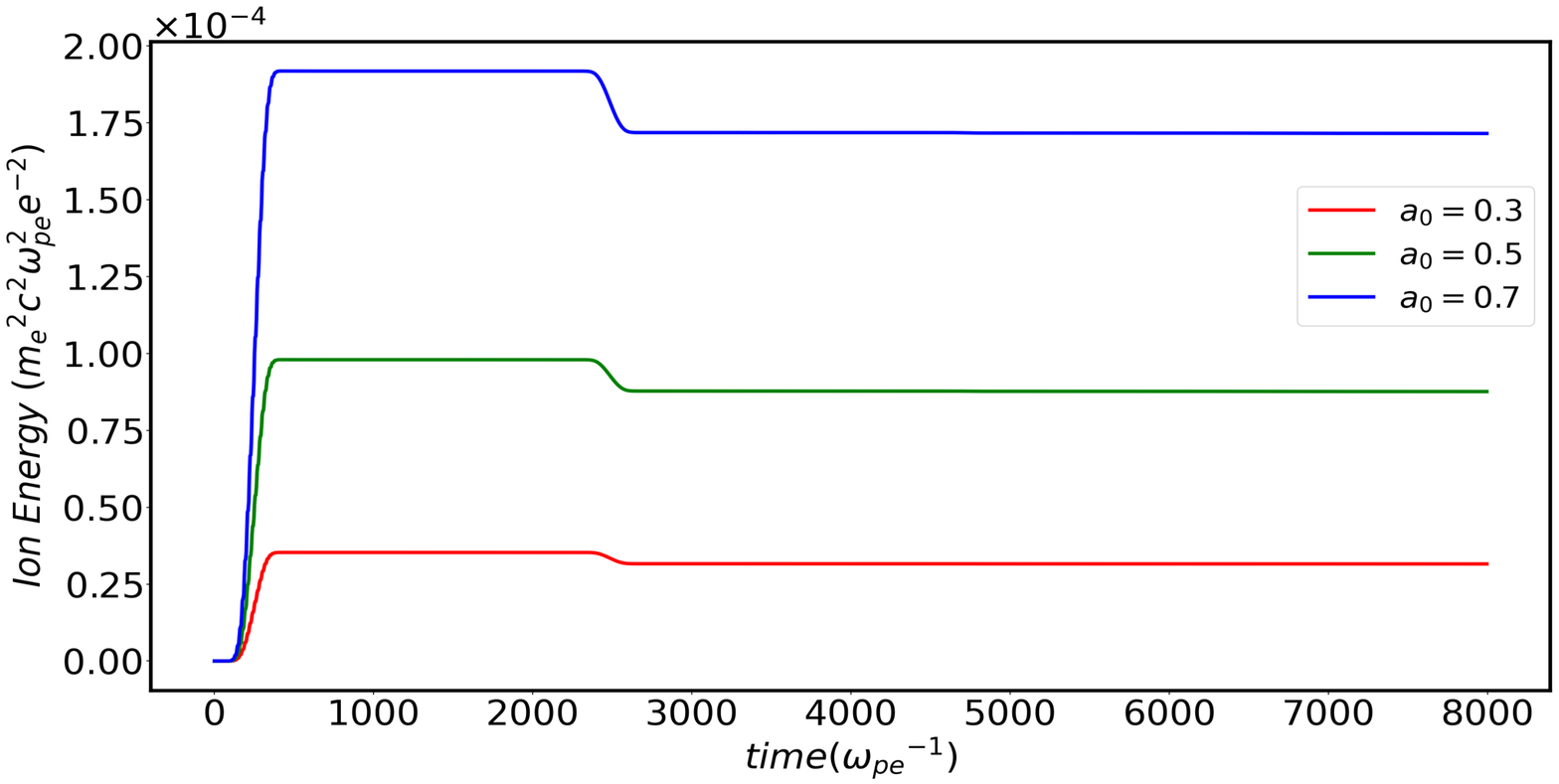}
  \caption{Comparison of energy absorbed by ions as a function of laser intensity in RL mode geometry for Case(II).}
\label{figure13}
\end{figure}

When the laser is incident on the sharp plasma boundary, a part of it gets reflected, and the other part gets transmitted. The transmission occurs only if   the frequency lies in the pass band of either $R$  or $L$ mode. The choice of magnetic fields is such that for $\omega_{ci} = 0.12\omega_{pe}$ the laser frequency ($\omega = 0.2\omega_{pe}$) lies in the pass band of $R$ wave, but the stop band of $L$ wave. We refer to this as Case(I). 
For the other two values of magnetic field and hence the gyrofrequencies, viz., $\omega_{ci} = 0.235\omega_{pe}$ (Case(II)) and $0.4\omega_{pe}$ (Case(III)) the laser frequency lies in the pass band of both modes. It is evident that $\omega_{0.235}$ is closer to the laser frequency. It is also important to note that while we have chosen the incident pulse to have linear polarization, as soon as the pulse enters the plasma, it breaks into the left and right-hand circularly polarized waves and propagates with their respective  group velocities. The group speed being different, the incident pulse separates between left and right circularly polarised pulses spatially \cite{goswami2021ponderomotive}. If the frequency lies in the stop band of one of the waves, the pulse corresponding to that polarization does not propagate inside the plasma.

In Fig.(\ref{figure7}) we show the evolution of the electrostatic energy, electromagnetic energy, and ion kinetic energy in subplots (a), (b), and (c), respectively, for all the three chosen values of the magnetic fields. 
It can be observed from the figure that the electrostatic and particle energy increases with time, whereas the electromagnetic energy associated with the incident wave decreases. The  sharp dips in the electromagnetic energy evolution indicate the time when either the  reflected wave  (from the plasma surface) or the transmitted wave moves  outside the simulation box. Even after the laser electromagnetic wave has left the simulation box,  both  electrostatic and kinetic  energies have finite values  in the medium.  This illustrates the  irreversible nature of the coupling of laser energy to the plasma medium.  The other thing to note is that the energy acquired by the ions is maximum for  Case(II), for which the laser frequency is closest to the ion gyro-frequency and is in the pass band of both  $L$ and $R$ waves. For Case(I) with $\omega_{ci} =0.12\omega_{pe}$, the energy acquired by the ions is the lowest. This is so as the propagating mode, in this case, has right-hand circular polarization, which is opposite to the direction of ion gyro-motion.  Even though the laser frequency $\omega =0.2\omega_{pe}$ is much less than the ion gyro-frequency in Case(III) 
 as compared to Case(I), the laser energy absorption by ions here is comparatively higher. This is so because in Case(III), the medium permits both $L$ and $R$ modes. The energy absorption in ions is more when the electric field rotation occurs  in the same direction as its gyro-motion.  In section \ref{sec5}, we study this effect in detail  by choosing both kinds of circular polarization  for the incident  electromagnetic wave.  It should, however, be noted that in  Case(III),   the electrostatic energy generation is negligible even though the ions acquire energy.   This suggests that the energy gain of  ions does not occur by any   charge separation mechanism for this  particular case. In fact,  the rotation of the electric field vector corresponding to $L$ mode being in the same direction as the ion gyro-rotation seems to be responsible for 
 ion energy gain. 

We next provide a comparison of the time evolution of  kinetic energy of ions and electrons for Case(II) in Fig.(\ref{Fig:RLmode-el+ion-compare}).  It  can be observed that energy absorbed by ions is much higher as compared to electrons for this case.

We have also carried out simulations spanning  an entire range of magnetic fields with shorter intervals.  This range  corresponds to having    $\omega_{ci}$  from $0.1\omega_{pe}$ to $0.6\omega_{pe}$. The laser frequency was  kept at   $\omega_{l}=0.2\omega_{pe}$. For $\omega_{ci}<0.2\omega_{pe}$, laser lies in the stop-band of L-mode and pass-band of R-mode, as shown in the dispersion curves in subplots (a) and (b) of Fig.(\ref{figure5}), respectively. Thus, as expected, only R-mode is generated inside the plasma. However, for  $\omega_{ci}>0.2\omega_{pe}$ laser lies in the passband of both L and R-mode, as shown in the dispersion curves in subplots (c)-(f) of Fig.(\ref{figure5}), respectively. Thus, both R and L-mode will be generated inside the plasma for the choice between this second range. At  $\omega_{ci}=0.2\omega_{pe}$, laser frequency matches exactly with the ion gyration frequency. The variation of ion energy and electrostatic energy with respect to the applied magnetic field in the plasma has been shown in the subplots (a) and (b) of Fig.(\ref{figure6}), respectively. 

It is clear from the plots that the maximum energy absorption occurs when the laser frequency is below the ion gyro-frequency, i.e., below ion cyclotron resonance. This is understood by realizing that when the laser frequency is equal to or higher than the ion cyclotron frequency, then only the right-hand circularly polarised wave propagates in the plasma for which the electric field rotates in the direction opposite to that of ion gyro-rotation.  On the other hand, when the laser frequency is lower than the ion gyro-frequency, it is in the pass-band of both $L $ and $R $ modes. As the ion  gyro-rotation is in the same direction as the electric field rotation for the L wave, resonance absorption occurs.   The most efficient coupling, therefore, occurs not at the resonance frequency but when the laser frequency is slightly lower than the ion gyro frequency. It is interesting to note that the electrostatic energy inside the plasma also  increases but suffers a sharp dip after the resonance. In fact, it vanishes to zero below the optimum  magnetic field required for the efficient absorption of energy by ions. This is understood by realizing that at laser frequencies lower than ion gyro-frequency, both ions and electrons exhibit a magnetized response. This is a regime where the drifts associated with the two species are similar, resulting in no charge separation. 
The ion energy absorption in this regime is guided by the resonance process and not the process discussed in 
(\cite{goswami2021ponderomotive}). 

We now study the possible spread of the energy acquired by the ions for all of these cases.
We show the ion counts as a function of  momentum  in both the propagation direction $ \hat{x}$ and in the perpendicular direction $\hat{y} $ for the three different cases of magnetic field strength described earlier. The  plots have been shown at two different times of  $t = 2000$ and $7800$ in Fig.(\ref{figure19}) and Fig.(\ref{figure8}), respectively. While at $t=2000$, the laser pulse is still inside the system at $t = 7800$, both the reflected and the transmitted part of the incident laser pulse have left the simulation box. It is interesting to note that for Case(I),  where only the $R$ mode 
is permissible in the system for which the electric field rotation is in a direction opposite to the ion gyro-rotation, the ions retain  momenta after the laser pulse has left  only along the propagation direction $\hat{x}$.  It should be noted that for this case, the electrostatic energy is finite, and the ions gain energy by the electrostatic field that gets generated due to the difference between the ponderomotive force experienced by the electron and ions by the laser pulse. 
The perpendicular momenta in ions remain finite only for the duration the laser pulse remains inside the system. 
 For Case(II), for which both $R$ and $L $ modes propagate inside the plasma, the momentum spread is observed in both the propagation and the perpendicular directions even after the pulse has left.  In this case, the electrostatic energy is negligible, suggesting that the difference between the ponderomotive pressure felt by the two species at this higher magnetic field is incapable of creating an electrostatic field. The ions here gain energy due to the ion gyro-rotation resonating with the electric field rotation of the $L$ wave propagating inside the medium. For  Case(III), both components of momenta are found to be insignificant. This is so as the laser frequency is away from the resonance, and the magnetic field is even higher compared to that of Case(II) for the ponderomotive force difference between the two species to generate an electrostatic field. 

 When the laser intensity is increased, as expected, the ion energy absorption also increases.  This has been depicted in the plot of Fig.(\ref{figure13}) where  red, green, and blue curves correspond to the values of  $a_{0}=0.3$, $0.5$, and $0.7$, respectively.

 \begin{figure}
  \centering
  \includegraphics[height=2.5in,width=6.0in]{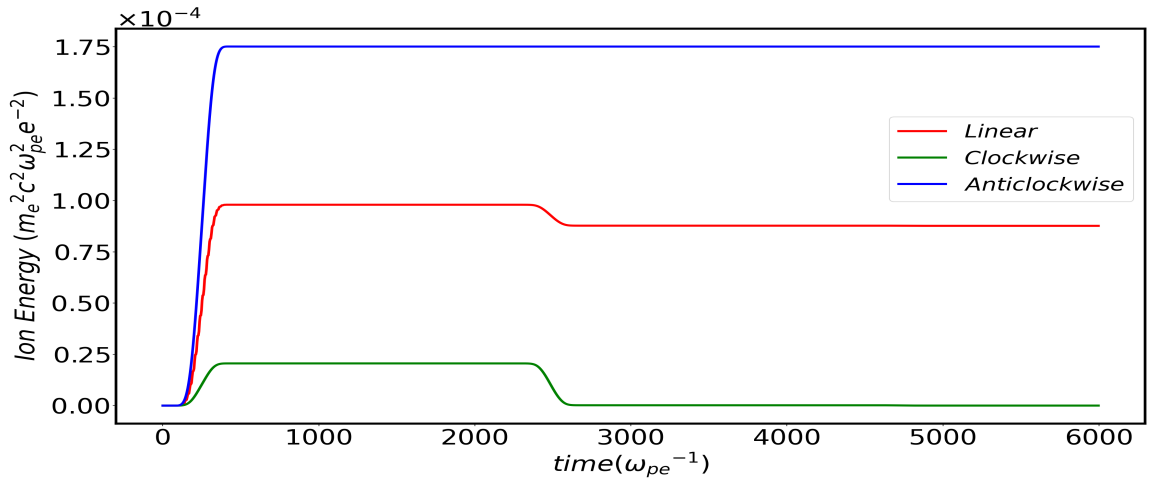}
  \caption{Energy absorbed by ions when the laser pulse is linearly, clockwise, and anticlockwise polarized in R-L Mode geometry.}
\label{Fig:RLmode-el+ion-with-polarization}
\end{figure}  

\begin{figure}
  \centering
  \includegraphics[height=2.5in,width=6.0in]{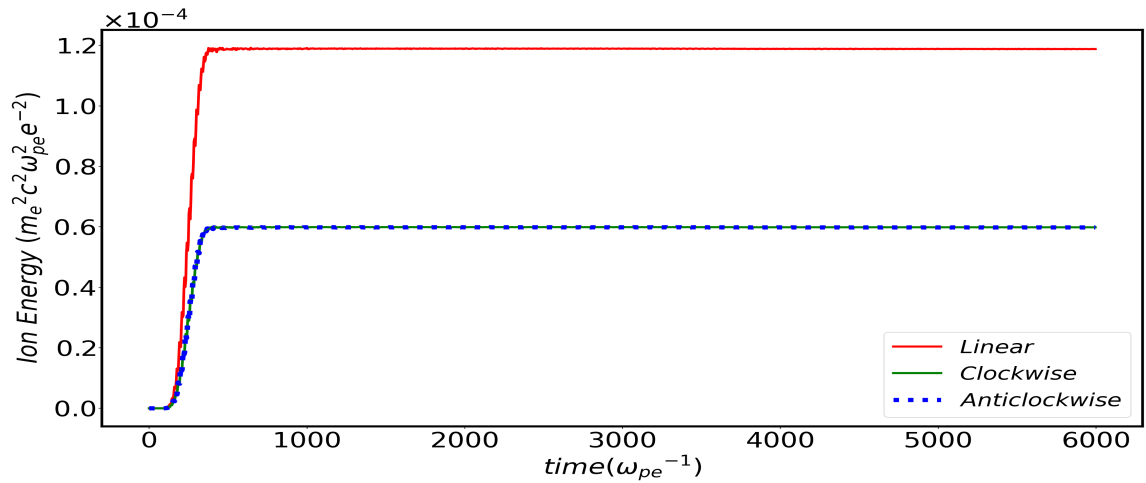}
  \caption{Energy absorbed by ions when the laser pulse is linearly, clockwise, and anticlockwise polarized in X-mode geometry.}
\label{Fig:Xmode-ion-polarization}
\end{figure}

\section{Role of Laser polarization\label{sec5}}
 In this section, we  study the role of incident laser polarization on ion energy absorption for both cases of RL and X mode configurations.  The  magnetic field is applied parallel to the propagation direction in the RL mode geometry. The studies here confirm that  the ions  gain maximum energy only when the incident laser is left-hand circularly polarised, i.e., the  electric field rotates in the same direction as the  ion gyro-frequency see Fig.(\ref{Fig:RLmode-el+ion-with-polarization}). For the  linearly polarised laser,  the energy gain is intermediate as it is  a combination of left and right circular polarised waves. 

For the X mode case, the absorption is maximum for the case of linear polarization, as shown in Fig.(\ref{Fig:Xmode-ion-polarization}). This is so because the  circularly polarised laser is a combination of two linear polarizations. One of which will be in the X mode configuration, while the other will be an ordinary mode for which the plasma is overdense.

\section{Conclusion\label{sec7}}
The manuscript discusses the possibility and mechanism of energy absorption by ions in the context of laser interacting with magnetized plasma. It is shown that  by restricting the dynamics of electrons with an externally applied magnetic field the laser energy can be channelized to the heavier ion species.  In addition, the  magnetic field even for overdense plasma  provides frequency bands  at which the propagation of the laser field is permitted inside the plasma. 

Both orientations of the externally applied magnetic field have been considered in these  studies, namely the X mode geometry and the R-L mode geometry. In X mode geometry the external magnetic field is oriented along the direction of the laser magnetic field and in the RL mode, it is along the laser propagation direction. In  the X mode configuration, the resonant excitation of the electrostatic lower hybrid mode is responsible for the preferential energy transfer to ions. However, the most efficient energy transfer is observed when the laser frequency is slightly lower and off-resonant. This permits the laser to fall in the pass band of the dispersion curve and also to remain close to the resonant condition.   

In the RL mode, on the other hand, two distinct mechanisms of laser energy absorption are observed. The first case is similar to the one observed by \cite{goswami2021ponderomotive}, wherein the difference in the ponderomotive force experienced by the two species produces  an electrostatic mode which then transfers energy to particles. In this case, both the electrostatic field and particle kinetic energy are observed to be finite. In addition to this phenomenon, we also observe an additional mechanism wherein the ions acquire energy directly by a resonant phenomenon. This happens when the  laser frequency matches with the ion - gyrofrequency for which the L mode resonantly couples with  the ion gyration. 
A detailed description has been provided for all these cases. The dependence on the magnetic field, laser polarization, and laser intensity has also been carried out in this work.

\section*{Acknowledgements}
The authors would like to acknowledge the OSIRIS Consortium, consisting of UCLA and IST (Lisbon, Portugal) for providing access to the OSIRIS-4.0 framework which is the work supported by the NSF ACI-1339893. AD would like to acknowledge her J C Bose fellowship grant of AD(JCB/2017/000055) and CRG/2018/000624 grant of DST. The authors thank IIT Delhi HPC facility for computational resources. RJ thanks the Council for Scientific and Industrial Research(Grant no.-09/086(1448)/2020-EMR-I) for funding the research.

\section*{References}
\bibliographystyle{unsrt}
\bibliography{Absorption.bib}

\end{document}